\documentclass{article}
\usepackage{epsfig,psfrag}
\begin{document}

\title{Microwave induced resonant backscattering in a one-dimensional mesoscopic channel}
\author{S.~Blom and L.Y.~Gorelik\\ \small{Department of Applied Physics, Chalmers University of Technology}\\\small{and G\"oteborg University, SE-412 96 G\"oteborg, Sweden}}
\date{November 22, 2000}
\maketitle

\begin{abstract}
We have studied a one-dimensional channel with a wider, straight region irradiated by an external electromagnetic field. In this system the interplay between interference effects and resonance phenomena manifests itself and provides a new behavior of conduction. We show that the transport properties of the system can be affected by changing frequency and the amplitude of the external field.  
We also show that for certain combinations of these parameters electron transport through the system is complitely blocked.
\end{abstract}

\section{Introduction}
Mesoscopic Physics, or mesoscopics, deals with systems that are large compared to the microscopic (atomic) scale but small compared to the macroscopic scale where normal Boltzmann transport theory is applicable for transport studies. Within the field of mesoscopics one major feature of interest is quantum interference phenomena, such as the Aharonov-Bohm effect~\cite{Beenakker} and the phenomena of universal conductance fluctuations~\cite{Ferry}. Different quantum mechanical waves (or different parts of a split wave) can (re)combine and interfere. Interference effects become important when the size of the system is small compared to the phase breaking length.
 
Another feature in systems of small size on the scale of the electron Fermi wavelength is that the motion of the electrons will be quantized into energy levels. This is the case in a narrow channel electrostatically formed by a split-gate on top of a heterostructure as well as in a nanowire. The quantization of the transverse motion of the electrons into one-dimensional modes is the cause of the quantized conductance seen in quantum point contacts (QPC's)~\cite{Wees,Wharam} and nanowires~\cite{Pascual}.

In a system with strong space quantization resonant interaction with an external electromagnetic field causes coupling between electron-modes. In this way transport properties may be strongly affected\cite{Feng,Blom,Tageman,Gorelik94,Gorelik97}.

In this paper we have studied a system where the interplay between interference effects and non-linear resonance phenomena manifests itself. We show that this interplay provides a new behavior of conduction.

\section{Model}
\begin{figure}
\begin{center}
\psfrag{a}[cc][cc]{$a$}
\psfrag{b}[tc][cc]{$b$}
\psfrag{c}[cc][cc]{$c$}
\psfrag{ct}[cc][cc]{$\tilde{c}$}
\psfrag{g}[cc][cc]{$\gamma$}
\psfrag{gt}[cc][cc]{$\tilde{\gamma}$}
\psfrag{al}[cc][cc]{$\alpha$}
\psfrag{be}[tc][cc]{$\beta$}
\psfrag{L}[cc][cc]{$L$}
\psfrag{EM}[Bc][cc]{EM-field}
\psfrag{x}[Bc][cc]{$x$}
\psfrag{y}[cc][cc]{$y$}
\psfrag{dx}[cc][cc]{$d(x)$}
\epsfig{file=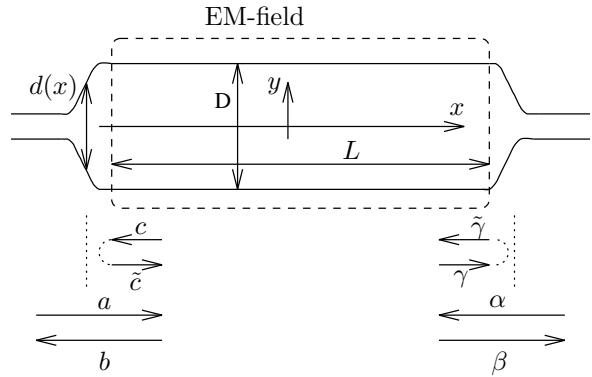, width=77mm}
\caption{We consider a channel constructed in a Si inversion layer 2DEG with a hard wall confining potential. There is a wider region with two open modes connected to more narrow channels, with only one mode, by an adiabatic QPC. In each mode electrons can travel in both directions. The upper mode is reflected in the QPC, ``trapped'' inside the wider region. The probability amplitudes corresponding to electrons in the upper mode are denoted $c$, $\tilde{c}$, $\gamma$ and $\tilde{\gamma}$. In the narrow ends of the channel there is only one open mode, the corresponding probability amplitudes are $a$, $b$, $\alpha$ and $\beta$ as shown in the figure. In the wider region we assume an external electromagnetic field that couples the two modes.\label{fig:sys}}
\end{center}
\end{figure}

 The system under study is a one-dimensional channel in a two-dimensional electron gas (a 2DEG) with a wider, straight, region irradiated by an external magnetic field, schematically depicted in figure~\ref{fig:sys}. Such a channel can be constructed e.g. in a 2DEG in a Si inversion layer using a split gate technique. We will consider the case when the system has geometry and field parameters such that only one mode in the narrow parts, connected to the leads, and two modes in the wider region, $-L/2<x<L/2$, are involved in the electron transport. The upper mode is reflected in the QPC's connecting the wider part to the more narrow parts of the channel. The probability amplitudes corresponding to electrons in different modes and traveling in different directions are denoted as shown in figure~\ref{fig:sys}.  

In the wider region there is an external electromagnetic field applied with amplitude $\varepsilon_\omega$ and frequency $\omega$ polarized in the $y$-direction. Such a field is described by the vectorpotential 
\begin{equation}
{\bf A}(x,t)=\frac{1}{\omega}\varepsilon_\omega\cos(\omega t){\bf e_y}
\end{equation}
The electromagnetic field vanishes in an unspecified way outside the wider region. The appropriate Schr\"odinger equation for an electron in the field region, $-L/2<x<L/2$, is 
\begin{equation}
i\hbar\frac{\partial}{\partial t}\Psi(x,y,t)=\left[\frac{1}{2m^*}\left[\hat{\bf p}+e{\bf A}(x,t)\right]^2+U(x,y)\right]\Psi(x,y,t)
\end{equation}

\section{Electron dynamics}
We consider a channel with a width, $d(x)$, which varies smoothly on the scale of the Fermi wavelength, hence the adiabatic approximation can be used, i.e. the transverse and longitudinal motion can be considered separately in the absence of the external field. The wavefuntion in the field region is~\cite{Tageman}
\begin{eqnarray}
& &\Psi(x,y,t) =\sum_{n\delta}\Psi_n^\delta(x,t)\Phi_n[y,d(x)]\\
& &\Psi_n^\delta(x,t)=\varphi_1^\delta(x)\frac{1}{\sqrt{v_1}}e^{i\delta{\mathcal S}}e^{-i[E+\hbar\omega(n-1)]t/\hbar}\\
& &{\mathcal S}=[S_{1,E}(x)+S_{2,E+\hbar\omega}(x)]/2\hbar\\
& & S_{n,E}(x)=\int_o^xdx'\sqrt{2m^*[E-E_n(x')]}\\
& & v_{n,E}=\frac{1}{m^*}\frac{\partial S_{n,E}(x)}{\partial x}
\end{eqnarray}
where the superscript $\delta$ indicates the two different directions the electron can travel in in each mode, left to right ($\delta=+1$) or right to left ($\delta=-1$). Specifically we consider a hard wall confining potential. Then each mode is described by
\begin{eqnarray}
&&\Phi_n(y,d(x))=\sqrt{\frac{2}{d(x)}}\sin\left[\frac{n\pi}{d(x)}\left(y+\frac{d(x)}{2}\right)\right]\\
&&E_n=\frac{\hbar^2}{2m^*}\frac{n^2\pi^2}{d^2(x)}
\end{eqnarray}
We will consider only two modes, $n=1,2$. Then the mode population amplitudes $\varphi_i^\delta(x)$ satisfy
\begin{equation}
i\frac{\partial}{\partial x}\overrightarrow{\varphi}^\delta(x)-P_E(x){\bf \sigma}_z\overrightarrow{\varphi}^\delta(x)+\Lambda_\omega{\bf \sigma}_y\overrightarrow{\varphi}^\delta(x)=0\label{eq:modepop}
\end{equation}
where we have neglected second order derivatives of $\varphi_i(x)$ and where ${\bf \sigma}_y$ and ${\bf \sigma}_z$ are Pauli matrices and
\begin{equation}
\overrightarrow{\varphi}^\delta(x)=\left[\begin{array}{c}\varphi_2^\delta(x)\\\varphi_1^\delta(x)\end{array}\right]
\end{equation}

\begin{eqnarray}
\Lambda_\omega & = & \frac{V_\omega}{\hbar\sqrt{v_{1,E}(x)v_{2,E+\hbar\omega}(x)}}\\
V_\omega & = & -\frac{4}{3}\frac{\hbar e \varepsilon_\omega}{m^*\omega d(x)}\\
P_E & = & \frac{S'_{1,E}(x)-S'_{2,E+\hbar\omega}(x)}{2\hbar}
\end{eqnarray}

The conditions for it to be reasonable to consider only two modes is that the width $d(x)$ and the frequency of the applied external field (see below) is such that for $d(x)=D$ (width of the wider region)
\begin{equation}
E_2(D)<\hbar\omega+E_F
\end{equation}
and such that for $d(x)=d_0$ (width of the narrow region)
\begin{equation}
E_1(d_0)<E_F<E_2(d_0))-\hbar\omega
\end{equation}

Since the channel has no $x$-dependence ($d(x)=D=$constant) in the irradiated region $P_E(x)={\mathcal{P}}_E$, independent of $x$, and we find the solutions
\begin{equation}
\overrightarrow{\varphi}^\delta_\pm(x)=\left(\begin{array}{c}\mp i\sqrt{1\mp\sin\lambda}\\\sqrt{1\pm\sin\lambda}\end{array}\right)e^{\pm i\delta K(E)x}\label{eq:sol}
\end{equation}
which are linearly independent and where
\begin{equation}
K(E)=\sqrt{{\mathcal{P}}^2_E+\Lambda^2_\omega}
\end{equation}
\begin{equation}
\sin\lambda=\frac{{\mathcal{P}}_E}{\sqrt{{\mathcal{P}}^2_E+\Lambda^2_\omega}}
\end{equation}
We take the total energy of the entering electron to be equal to the Fermi energy $E=E_F$.

\section{Transport properties}
\begin{figure}
\begin{center}
\psfrag{a}[cc][cc]{$a$}
\psfrag{b}[cc][cc]{$b$}
\psfrag{c}[cc][cc]{$c$}
\psfrag{ct}[cc][cc]{$\tilde{c}$}
\psfrag{g}[cc][cc]{$\gamma$}
\psfrag{gt}[cc][cc]{$\tilde{\gamma}$}
\psfrag{al}[cc][cc]{$\alpha$}
\psfrag{be}[cc][cc]{$\beta$}
\epsfig{file=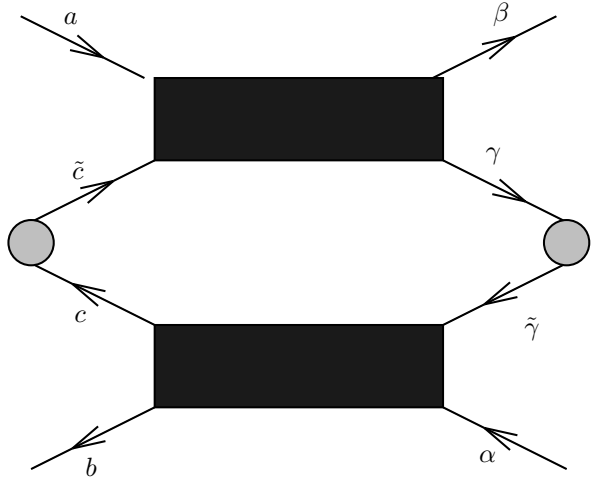, width=77mm}
\caption{A schematic view of our system. The upper black box corresponds to the coupling between the lower and upper modes inside the field region for electrons moving from left to right. The lower black box is for the coupling between electrons going in the opposite direction. The phaseshift that the electron suffers when reflected in the QPC is indicated by a gray circle.\label{fig:sch}}
\end{center}
\end{figure}

The conductance in a ballistic system is given by the transmission through the well-known Landauer formula. So, in order to describe the transport properties of our system we need to find the transmission and reflection,  i.e. the elements of the scattering matrix which connect the probability amplitudes in the two narrow ends, given by
\begin{equation}
\left(\begin{array}{c}b\\ \beta\end{array}\right)=\left(\begin{array}{cc} \rho & \tau\\ \tau & \rho\end{array}\right)\left(\begin{array}{c} a\\\alpha\end{array}\right)
\end{equation}
We can calculate these elements by noting that our system is similar to a general two barrier system. Transmission in our system corresponds to reflection in the two barrier problem, and vice versa. The similarity between our system and a general two barrier problem become transparent if we schematically describe our system as in figure~\ref{fig:sch}. The electromagnetic field only couples electrons moving in the same direction, hence we describe this coupling by two different boxes: one for electrons moving from left to right and one for electrons going in the other direction. If an electron comes out of a box in an upper mode ($c$ or $\gamma$) it will be reflected in the QPC and enter the other box. When reflected the phase of the electron will be shifted by $\pi/2$~\cite{LandauLifshitz}, in the figure this is indicated by a gray circle.

The scattering matrices describing the two different boxes can easily be found using equation~\ref{eq:modepop}. The matrix for motion from left to right is
\begin{equation}
\left(\begin{array}{c}\gamma\\ \beta\end{array}\right)=e^{i\kappa L}\left(\begin{array}{cc} s_{12} & s_{11}^*\\ s_{11}& -s_{12}\end{array}\right)\left(\begin{array}{c} a\\ ce^{-i\pi/2}\end{array}\right)\label{eq:sys1}
\end{equation}
and for motion from right to left
\begin{equation}
\left(\begin{array}{c}b\\ c\end{array}\right)=e^{i\kappa L}\left(\begin{array}{cc} -s_{12} & s_{11}\\ s_{11}^* & s_{12}\end{array}\right)\left(\begin{array}{c} \gamma e^{-i\pi/2} \\ \alpha \end{array}\right)\label{eq:sys2}
\end{equation}
The behavior in the field-region is described by the solution above (eq.~\ref{eq:sol}) which gives us the matrix elements
\begin{eqnarray}
s_{12} & = & |\cos\lambda|\sin(KL)\label{eq:s12}\\
s_{11} & = & \cos(KL)+i\sin(KL)\sin\lambda\label{eq:s11}
\end{eqnarray}
The phase shift due to traveling through the irradiated region, $e^{i\kappa L}$ is given by
\begin{eqnarray}
\kappa & = & \frac{1}{2\hbar}\left(\sqrt{2m^*(E-E_1)}+\sqrt{2m^*(E+\hbar\omega-E_2)}\right)
\end{eqnarray}
We see that $s_{11}$, $s_{12}$ and $\kappa$ and hence the transport properties of the system depend on the amplitude and  frequency of the external field as well as on the length of the irradiated region.

\subsection{Reflection and transmission}
Instead of solving the system of equations (eq.~\ref{eq:sys1} and~\ref{eq:sys2}) we can use the result from the two barrier problem to simply write down the elements of the scattering matrix for our system:
\begin{eqnarray}
\tau & = & e^{i\kappa L}s_{11}+\frac{e^{3i\kappa L}s_{11}^*(s_{12})^2}{1+e^{2i\kappa L}(s_{11}^*)^2}\\
\rho & = & \frac{ie^{2i\kappa L}(s_{12})^2}{1+e^{2i\kappa L}(s_{11}^*)^2}\label{eq:rho}
\end{eqnarray}

We see that the total transmission, $T=\tau\tau^*$, is perfect (equal to unity) whenever $\rho\rho^*=0$, i.e when, according to eq.~\ref{eq:rho}, $s_{12} ^2=0$, and since $cos\lambda\neq 0$, we find the following condition for perfect transmission, see eq.~\ref{eq:s12},
\begin{equation}
KL=L\sqrt{{\mathcal P}_E^2+\Lambda_\omega^2}=m\pi\label{eq:analytic}
\end{equation}
where $m$ is an integer. Each integer will give a line of perfect transmission when varying the amplitude $\varepsilon_\omega$ and the frequency $\omega$ of the electromagnetic field. 

Perfect reflection (zero transmission) will occur when $|\rho|=1$. Due to the unitarity condition for scattering matrices $|s_{12}|^2=1-|s_{11}|^2$ so that we can rewrite $\rho$ as
\begin{equation}
\rho=\frac{ie^{2i\kappa L}(1-|s_{11}|^2)}{1+e^{2i(\kappa L-\chi)}|s_{11}|^2}\end{equation}
where $\chi$ is the argument of $s_{11}$,
\begin{equation}
\chi=\arctan\frac{\sin(KL)\sin\lambda}{\cos(KL)}\label{eq:chi}
\end{equation}
The condition for perfect reflection is that $e^{2i(\kappa L-\chi)}=-1$, i.e. that
\begin{equation}
2(\kappa L-\chi)=(2m+1)\pi \label{eq:perref}
\end{equation}
where $m$ is an integer. This condition will be fulfilled for certain choices of parameters (amplitude and frequency) giving lines of perfect reflection in the transmission landscape.

\section{Results and Discussion}
In the numerical calculations we have used parameters corresponding to a Si inversion layer: Fermi energy $E_F=2$ meV and effective mass $m^*=0.19m_e$; and $D=60$ nm.

\begin{figure}
\begin{center}
\psfrag{amp}[tc][bc]{$\varepsilon_\omega$ (V/mm)}
\psfrag{freq}[bc][tc]{$\omega$ $(\Delta E/\hbar)$}
\psfrag{trans}{ }
\epsfig{file=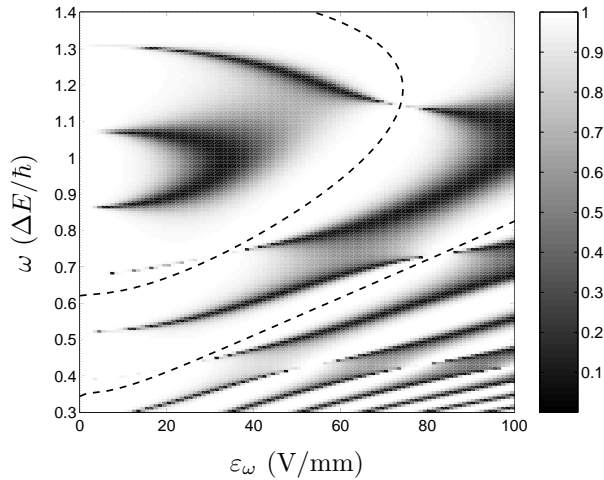, width=77mm}
\caption{Total transmission as a function of the amplitude, $\varepsilon_\omega$, and the frequency, $\omega$,(in units of $\Delta E/\hbar$ where $\Delta E=E_2-E_1$) of the external electromagnetic field. The length of the channel is in this case $L=16.37\frac{\pi}{2}/\kappa\approx 302$ nm. The dashed lines correspond to lines where the transmission is unity, given by eq.~\ref{eq:analytic}. For clarity only two of these lines are shown. \label{fig:trans_asym}}
\end{center}
\end{figure}

\begin{figure}
\begin{center}
\psfrag{amp}[tc][bc]{$\varepsilon_\omega$ (V/mm)}
\psfrag{freq}[bc][tc]{$\omega$ $(\Delta E/\hbar)$}
\psfrag{trans}{ }
\epsfig{file=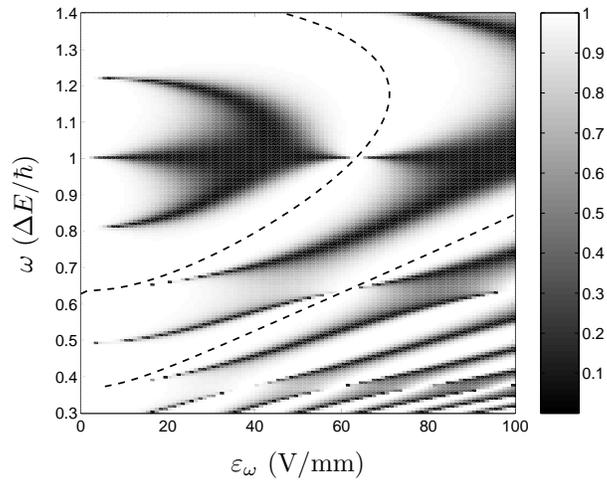, width=77mm}
\caption{Total transmission as a function of the amplitude, $\varepsilon_\omega$, and the frequency, $\omega$, (in units of $\Delta E/\hbar$ where $\Delta E=E_2-E_1$) of the external electromagnetic field. The length of the channel is in this case  $L=17\frac{\pi}{2}/\kappa\approx 314$ nm. For this length the condition for perfect reflection is fulfilled for all amplitudes at resonance ($\hbar\omega=\Delta E$). The dashed lines correspond to lines where the transmission is unity, given by eq.~\ref{eq:analytic}. For clarity only two of these lines are shown.\label{fig:trans_pr}}
\end{center}
\end{figure}

The total transmission of the considered system as a function of the amplitude and the frequency (in units of $\Delta E/\hbar$ where $\Delta E=E_2-E_1$) of the external electromagnetic field is shown in figures~\ref{fig:trans_asym} and~\ref{fig:trans_pr}. The dashed lines correspond to lines where the transmission is perfect (unity), given by eq.~\ref{eq:analytic}. For clarity only two of these lines are shown. 

There are two different mechanisms that can give perfect reflection: Rabi oscillations and interference effects. The first is when all electrons entering in the lower mode are in the upper mode at the other end of the channel and hence are reflected. Once they reach the first end again they will be back in the lower mode and can exit the field region, hence in this case the transmission equals zero. Looking at the scattering matrix element $s_{12}$ (eq.~\ref{eq:s12}) we see that this ``Rabi reflection'' can only happen when $sin\lambda=0$ (i.e. in resonance, $\omega=\Delta E/\hbar$) and the amplitude is such that the length of the field region corresponds perfectly to $\frac{(m+1/2)}{2}$ Rabi wavelengths, where $m$ is an integer.

The condition for perfect reflection due to interference effects, eq.~\ref{eq:perref}, can be fulfilled also out of resonance. When this conditions is fulfilled it means that summing up the probability amplitudes of an electron going through the system after having bounced back and forth multiple times gives zero.

 At  resonance the condition for perfect reflection due to interference becomes
\begin{equation}
\kappa L=(2m+1)\frac{\pi}{2}
\end{equation}
(see eq.~\ref{eq:chi} and~\ref{eq:perref}). In figure~\ref{fig:trans_pr} the length of the channel corresponds to  $\kappa L=301\frac{\pi}{2}$ at resonance. This means that the condition for perfect reflection due to interference is fulfilled at $\hbar\omega=\Delta E$ for any amplitude $\varepsilon_\omega$.

Perfect transmission in this system is a Rabi phenomena. It happens whenever (in resonance as  well as out of resonance) all the electrons entering on one side of the channel in the lower mode are all in the lower mode at the other end of the channel, from which they then simply can exit. This will be the case when the amplitude of the electromagnetic field is such that the length of the field region corresponds perfectly to an integer number of half Rabi wavelengths.

At resonance conduction is a periodic function of $\varepsilon_\omega$ with period
\begin{equation}
\Pi_\varepsilon=\lambda_R\frac{\varepsilon_\omega}{L}=\frac{3\pi m^*\omega d v_1}{4Le}
\end{equation}
where $\lambda_R$ is the Rabi wavelength. For the parameters in figure~\ref{fig:trans_asym} this equals 66 V/mm.

\begin{figure}
\begin{center}
\psfrag{trans}[bc][tc]{transmission}
\psfrag{freq}[cc][cc]{$\omega (\Delta E/\hbar)$}
\psfrag{trans2}[bc][tc]{transmission}
\psfrag{freq2}[cc][cc]{$\omega (\Delta E/\hbar)$}
\epsfig{file=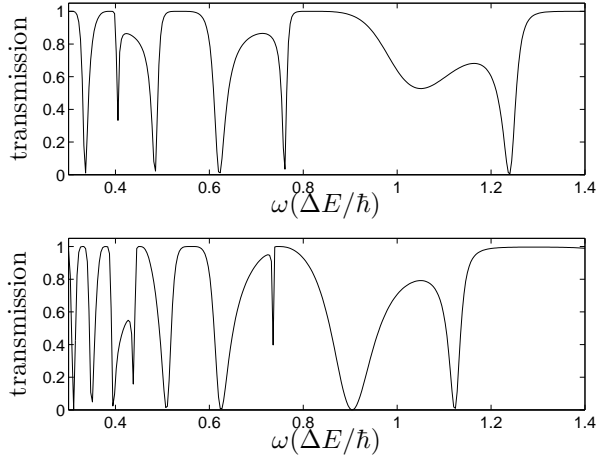, width=77mm}
\caption{Total transmission as a function of frequency for fixed field amplitude. In the upper picture $\varepsilon_\omega=45$ V/mm and in the lower $\varepsilon_\omega=85$ V/mm. Some of the dips are so narrow that they are hard to resolve.\label{fig:cross_freq}}
\end{center}
\end{figure}

\begin{figure}
\begin{center}
\psfrag{trans}[bc][tc]{transmission}
\psfrag{amp}[cc][cc]{$\varepsilon_\omega$ (V/mm)}
\psfrag{trans2}[bc][tc]{transmission}
\psfrag{amp2}[cc][cc]{$\varepsilon_\omega$ (V/mm)}
\epsfig{file=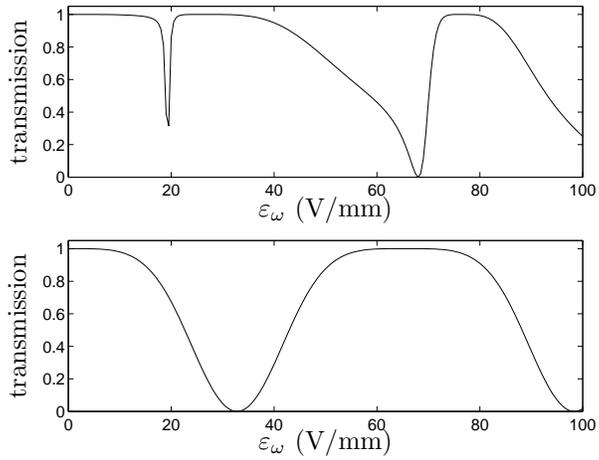, width=77mm}
\caption{Total transmission as a function of amplitude, $\varepsilon_\omega$, for fixed frequency. In the upper picture $\hbar\omega/\Delta E=0.7$ and in the lower picture $\hbar\omega/\Delta E=1$. In the lower picture we see that the transmission oscillates with half the Rabi wavelength as discussed in the text.\label{fig:cross_amp}}
\end{center}
\end{figure}

The transmission as a function of frequency (keeping $\varepsilon_\omega$ constant) is shown in figure~\ref{fig:cross_freq} and  as a function of amplitude for constant $\omega$ in figure~\ref{fig:cross_amp}. The length of the channel in these graphs are the same as in figure~\ref{fig:trans_asym}. 

When the amplitude of the external field becomes smaller the system is very sensitive and the transmission dips of perfect reflection become very narrow, the white lines in figures~\ref{fig:trans_asym}-\ref{fig:trans_pr} become thinner. In fact they become hard to resolve.

One limitation for observating the discussed phenomena in an experiment is 
that the temperature will smear out the sharp conductance dips associated 
with the resonant backscattering process, due to the strong energy dependence 
of the phase $2\kappa L$. In order to have a sharp dip, $2\kappa L$ must vary 
significantly less than $2\pi$ within the temperature range. This gives us the 
following condition on the temperature
\begin{equation}
T<\frac{\hbar v_F}{k_B\pi L}\sim 0.5K
\end{equation}

Being a phase coherent phenomena by nature resonant backscattering effects 
can also be destroyed by inelastic relaxation of the photoexcited electrons. 
(Elastic intermode scattering would also destroy the coherence, but this 
effect is weak for the narrow channel we consider~\cite{Maao96}.) This process may 
be described by introducing an imaginary part of the phase $2\kappa L$, 
$\mbox{Im}\{2\kappa L\}=L/l_{in}$, where $l_{in}$ is the inelastic mean free path. The typical excess energy $\Delta E$ corresponds roughly to the energy spacing between 
the two transverse levels, which for our system is 19 K. We can use the experimental results by St\"oger et al.~\cite{Stoger} to roughly estimate the inelastic electron mean free path to be of the order of $1$ $\mu$m for $\Delta E\sim 20$ K. This is sufficiently long to make the effects discussed in this work observable.

In the crossing between perfect transmission and perfect reflection, the transmission is expected to be close to one since the perfect reflection is a resonent effect which in a realistic setup is supressed because of the reasons mentioned above.

\section{Conclusions}
We have shown that the transport properties of a one-dimensional channel with an irradiated widened region can be affected by changing the amplitude or frequency of the external electromagnetic field. This is due to a combination of resonance and interference effects. We have shown that for certain combinations of the external paramters $\omega$ (field frequency) and $\varepsilon_\omega$ (field amplitude) the system has perfect reflection and perfect transmission respectively.\vspace*{0.5cm}\\

Valuable discussions with Robert Shekhter and Anatoli Kadigrobov are greatfully acknowledged.
This work was done with support from the Swedish Natural Research Council and the Swedish Foundation for Strategic Research.


\begin{thebibliography}{99}
\bibitem{Beenakker}C. W. J. Beenakker and H. van Houten, in {\em Solid State Physics}, vol. 44, ed. by H. Ehrenreich and D. Turnbull, Academic (San Diego, 1991).

\bibitem{Ferry}David K. Ferry and Stephen M. Goodnick, {\em Transport in Nanostructures}, Cambridge University Press (Cambridge 1997).

\bibitem{Wees}B. J. van Wees, H. van Houten, C. W. J. Beenakker, J. G. Williamson, L. P. Kouwenhoven, D. van der Marel and C. T. Foxon, Phys. Rev. Lett. {\bf 60}, 848 (1988).

\bibitem{Wharam}D. A. Wharam, T. J. Thornton, R. Newbury, M. Pepper, H. Ahmed, J. E. F. Frost, D. G. Hasko, D. C. Peacock, D. A. Ritchie and G. A. C. Jones, J. Phys. C {\bf21}, L209 (1988).

\bibitem{Pascual}J. I. Pascual, J. M\'endez, J. G\'omez-Herrero, A. M. Bar\'o, N. Garcia and Vu Thien Binh, Phys. Rev. Lett. {\bf 71}, 1852 (1993).

\bibitem{Feng}Shechao Feng and Qing Hu, Phys. Rev. B {\bf 48}, 5354 (1993).

\bibitem{Blom}S. Blom, L. Y. Gorelik, M. Jonson, R. I. Shekhter, A. G. Scherbakov, E. N. Bogachek and U. Landman, Phys. Rev. B {\bf 58}, 16305 (1998).

\bibitem{Tageman}Ola Tageman, L. Y. Gorelik, R. I. Shekhter and M. Jonson, J. Appl. Phys {\bf 81}, 285, 1997.

\bibitem{Gorelik94}L. Y. Gorelik, Anna Grincwajg, V. Z. Kleiner, R. I. Shekhter and M. Jonson, Phys. Rev. Lett. {\bf 73}, 2260 (1994).

\bibitem{Gorelik97}L.Y. Gorelik, Frank A. Maa{\o}, R. I. Shekhter and M. Jonson, Phys. Rev. Lett. {\bf 78}, 3169 (1997).

\bibitem{LandauLifshitz}L. D. Landau and E. M. Lifshitz, {\em Quantum Mechanics} 3rd ed., Butterworth Heinemann (Oxford 1997), sec. 47.

\bibitem{Maao96}Frank A. Maa{\o} and L. Y. Gorelik, Phys. Rev. B {\bf 53}, 15885 (1996).

\bibitem{Stoger}G. St\"oger, G. Brunthaler, G. Bauer, K. Ismail, B. S. Meyerson, J. Lutz and F. Kuchar, Phys. Rev. B {\bf 49}, 10 417 (1994).

\end{thebibliography}
\end{document}